# Structured 3D-SVD: A Practical Framework for the Compression and Reconstruction of Biological Volumetric Images

Mario Aragonés Lozano, Oscar Romero and Antonio León

Communications Department, Universitat Politècnica de València, 46022 Valencia, Spain

**Abstract.** This work introduces Structured 3D-SVD as a practical framework for the reconstruction, compression, and analysis of biological volumetric data. Inspired by the logic of matrix singular value decomposition (SVD), the proposed approach represents third-order volumetric data in the spatial domain and supports progressive reconstruction through ordered quasi-singular coeffients. The experimental evaluation was carried out on two biological volumetric datasets: one full-volume scan of a fish and another of a brain. The results show that Structured 3D-SVD achieves reconstruction quality close to that of Tucker decomposition while requiring shorter computation times and outperforms canonical polyadic decomposition (CPD) in both accuracy and runtime. In addition, a progressive reconstruction analysis shows that relatively low truncation levels are sufficient to preserve the main volumetric structures, while higher truncation levels lead to more detailed reconstructions.

## 1   Introduction

### 1.1   Background and Motivation

The analysis, compression, and efficient representation of three-dimensional data are major challenges in a wide range of scientific and technological disciplines. In particular, applications in biological imaging such as volumetric microscopy, 3D animal morphology scanning, and anatomical imaging of fish and other specimens require the handling of large 3D datasets that must be stored efficiently while still allowing structural analysis and reconstruction.

Effective compression of such data requires identifying and exploiting spatial patterns across the three dimensions of the volume. In the two-dimensional case, the SVD is a widely used tool for compression, denoising, filtering, and dimensionality reduction. Its ability to organize information in a compact and ordered way has made it especially useful in image processing and related numerical applications.

The SVD of a matrix $A \in R^{m \times n}$ can be written as $A = U\Sigma V^T$, where $U$ and $V$ are orthogonal matrices and $\Sigma$ is diagonal with singular values arranged in decreasing order. This representation provides a natural mechanism for progressive approximation: by retaining only the rst $k$ singular values, one obtains a reduced representation that often preserves most of the relevant structure of the data.

In many applications, however, the data are volumetric and represented as a three-dimensional tensor $X \in R^{n1 \times n2 \times n3}$. It is therefore natural to seek SVD-inspired strategies in such contexts, especially when progressive reconstruction and compact representation are desired. Unlike matrices, tensors do not admit a single canonical notion of multiplication, diagonalization, or rank, which makes the transfer of matrix-based ideas to higher dimensions considerably more involved.

Several tensor decomposition techniques have been proposed for the analysis and compression of high-dimensional data. Classical methods such as Tucker and CPD were reviewed by Kolda and Bader [1], establishing the mathematical foundations for tensor approximations. A foundational multilinear generalization of matrix SVD is the HOSVD, introduced in [2], which extracts orthogonal factor matrices along the different tensor modes. Tucker decomposition represents a tensor through a smaller core and a set of factor matrices, while CPD expresses it as a sum of rank-one terms. The tensor singular value decomposition (t-SVD), introduced by Kilmer and Martin [3], transfers matrix SVD ideas to third-order tensors by operating in the Fourier domain. These approaches have led to important advances in tensor computation and approximation. However, they do not directly provide the type of simple spatial-domain reconstruction framework considered in this work, where the emphasis is on practical progressive reconstruction of volumetric data.

Beyond these classical decompositions, current tensor-based and data-driven approaches also include formats such as tensor train and tensor ring, as well as deep learning-based methods for volumetric representation and reconstruction.

Recent studies have further expanded the range of tensor-based techniques and applications. Burch et al. [4] highlighted the relevance of tensor models in biomedical imaging, including possible links to quantum computing frameworks. Zhang and Golub [5] discussed rank-one tensor approximation strategies, whereas Qi et al. [6] studied tensor eigenvalue formulations in more algebraic settings. General computational aspects of tensor methods are discussed in [7], and efficient tensor recovery techniques have also been proposed in machine learning contexts, including the anchor-based framework introduced in [8].

Tensor-based methods have also been applied to color video compression [9], imaging [10], and hyperspectral data analysis [11]. Additional approaches include patch-based tensor models for hyperspectral unmixing [12]; multichannel cell imaging [13]; Tucker compression for magnetic resonance imaging (MRI) [14]; tensor-train reduced models for parameterized partial differential equations [15]; and recent developments in optical coherence tomography (OCT) imaging [16], ultrasound tensor imaging [17], and phase-contrast tensor microscopy [18]. In biological and anatomical imaging, Gignac et al. [19] and the MorphoSource initiative [20] provide relevant context for volumetric specimen data such as those used in our experiments. Related overviews in the eld of biomedical imaging can also be found in [21, 22].

Other matrix-inspired tensor decompositions have also been explored. For example, ref. [23] introduces a tensor generalized Schur decomposition in the Fourier domain, with applications to generalized eigenvalue problems and related areas. In contrast, the present work does not rely on spectral transforms or generalized eigenvalue formulations but instead builds a representation directly in the spatial domain for progressive volumetric reconstruction.

The goal of this paper is to present Structured 3D-SVD as a practical framework for the reconstruction, compression, and analysis of biological volumetric data. Inspired by the logic of matrix SVD, the proposed method provides a representation of volumetric tensors that enables progressive approximation and reconstruction as the truncation level increases. The proposed quasi-singular coefficients are introduced to support progressive approximation and reconstruction of 3D volumetric data.

In this work, we refer to the proposed framework as Structured 3D-SVD, emphasizing the organization of third-order data for progressive truncation and reconstruction. From a computational viewpoint, the proposed construction is based on decompositions along the tensor modes and a core representation, with emphasis on practical coefficient organization and reconstruction behavior for 3D imaging data.

Structured 3D-SVD should be understood as a practical framework for progressive reconstruction of third-order volumetric data in the spatial domain, rather than as a new canonical tensor decomposition in the strict algebraic sense. While the proposed method shares with Tucker/HOSVD the use of modewise basis matrices and a reduced core tensor, its main contribution lies in the structured organization and interpretation of the representation for practical truncation and reconstruction. In particular, the method introduces ordered quasi-singular coefcients derived from the reduced core tensor and uses them to support progressive reconstruction and truncation analysis of volumetric biological data.

The proposed method is evaluated on biological volumetric datasets with different anatomical characteristics and levels of spatial complexity. Reconstruction quality is assessed using standard metrics such as peak signal-to-noise ratio (PSNR), mean squared error (MSE), relative reconstruction error, and percentage of energy retained (PER), used here as a practical cumulative indicator for truncation analysis, to evaluate the suitability of the method for compact representation and progressive reconstruction.

### 1.2 *Main Contributions*

The main contributions of this work are as follows:
- We present Structured 3D-SVD as a practical framework for volumetric tensor representation and progressive reconstruction in the spatial domain.
- We introduce an ordered coefficient organization based on quasi-singular coefficients derived from the reduced representation, which supports practical truncation analysis and progressive reconstruction.
- We demonstrate the practical usefulness of the proposed method on real biological volumetric datasets, showing reconstruction quality close to Tucker decomposition with reduced computation times and better results than CPD in both accuracy and runtime.

### 1.3 *Paper Organization*

The rest of the paper is organized as follows. Section 2 presents the tensor notation and basic concepts used throughout the paper. Section 3 presents the Structured 3D-SVD method and its computational procedure. Section 4 provides an experimental evaluation on volumetric biological datasets and discusses the results. Finally, Section 5 presents the conclusions and possible future work.

## 2 Tensor Notation and Basic Concepts

This section introduces the notation and basic concepts used in the proposed Structured 3D-SVD framework. The purpose is not to develop a general tensor theory but to establish the elementary tools required to describe the representation and reconstruction procedure.

Let $A, B \in R^{n_1 \times n_2 \times n_3}$ be two third-order tensors. Their inner product is deffned as

$$\langle A, B \rangle = \sum_{i=1}^{n_1} \sum_{j=1}^{n_2} \sum_{k=1}^{n_3} A_{ijk} B_{ijk}. \tag{1}$$

The associated Frobenius norm is given by

$$\| A \|_F = \sqrt{\langle A, A \rangle}. \tag{2}$$

This norm is used throughout the paper to measure reconstruction errors and approximation quality.

Given three vectors $u \in R^{n1}$, $v \in R^{n2}$, and $w \in R^{n3}$, their outer tensor product defines a rank-one tensor:

$$(u \otimes v \otimes w)_{jkm} = u_j v_k w_m. \tag{3}$$

This type of separable component is the basic building block used in the proposed representation.

In addition, the method makes use of the modewise unfoldings of a tensor. Given a tensor $X \in R^{n1 \times n2 \times n3}$, its unfoldings $X_{(1)}$, $X_{(2)}$, and $X_{(3)}$ are the matrix rearrangements obtained by organizing the tensor entries along modes 1, 2, and 3, respectively. These unfoldings make it possible to extract modewise factor matrices by standard matrix SVD.

Within the proposed framework, the reconstruction coefficients are organized in decreasing order and used to generate progressively more accurate approximations of the original volume.

## 3 Structured 3D-SVD Method

This section presents the proposed Structured 3D-SVD framework for volumetric data representation, practical truncation analysis, and progressive reconstruction of 3D volumes in the spatial domain. The method emphasizes ordered coefficient organization as the basis for progressive reconstruction. Table 1 summarizes the main conceptual and practical differences between Structured 3D-SVD and the reference tensor-based methods considered in this work.

**Table 1**: Conceptual and computational comparison between related tensor-based methods.

| Method | Domain | Ordered Scalar Coefficients | Progressive Reconstruction from One Computed Model | Main Practical Distinction/Limitation |
|---|---|---|---|---|
| Tucker/HOSVD | Spatial domain | No | Partial | Uses factor matrices and a core tensor but does not provide an explicitly ordered 1D coefficient organization for progressive truncation. |
| CPD | Spatial domain | Partial | Limited | Represents the tensor as a sum of rank-one terms but typically requires iterative fitting and is sensitive to rank selection and initialization. |
| Structured 3D-SVD | Spatial domain | Yes | Yes | Organizes the reduced representation through ordered quasi-singular coefficients for practical truncation and progressive reconstruction. |

## 3.1 Representation of a Volume

Given a real 3D tensor $X \in R^{n1 \times n2 \times n3}$, the proposed method builds a representation of the volumetric data from factor matrices along the tensor modes and associated ordered coefficients. This representation is designed to support progressive reconstruction: as the truncation level increases, the approximation becomes increasingly accurate.

In this framework, the volume is described through vectors $\mathbf{u}^{(i)} \in R^{n1}$, $\mathbf{v}^{(i)} \in R^{n2}$, and $\mathbf{w}^{(i)} \in R^{n3}$, together with coefficients $q\sigma_i$, which are referred to in this work as quasi-singular coefficients. These coefficients are used to organize their contribution in decreasing order, in a manner inspired by the role of singular values in matrix SVD.

For interpretation purposes, the structured representation can be expressed through the following ordered rank-one expansion:

$$X \approx \sum_{i=1}^{r} q\,\sigma_i\, u^{(i)} \otimes v^{(i)} \otimes w^{(i)}. \tag{4}$$

where $\otimes$ denotes the outer tensor product in three dimensions. Each term represents a separable volumetric component associated with the three modes of the tensor. This expansion is used here as an interpretive representation of the ordered components, while the actual truncated reconstructions are obtained from the reduced core tensor and the corresponding truncated factor matrices. If we denote as

$$\bar{X}_r = \sum_{i=1}^{r} q\,\sigma_i\, u^{(i)} \otimes v^{(i)} \otimes w^{(i)}, \tag{5}$$

the approximation associated with Equation (4), then its relative approximation error can be quantified as

$$\varepsilon_r = \frac{\parallel X - \bar{X}_r \parallel_F}{\parallel X \parallel_F}. \tag{6}$$

Since the vectors in each mode are orthonormal, the rank-one tensors

$$u(i) \otimes v(i) \otimes w(i)$$

are mutually orthonormal, and therefore,

$$\parallel X - \bar{X}_r \parallel_F^2 = \parallel X \parallel_F^2 - \sum_{i=1}^{r}(q\sigma_i)^2. \tag{7}$$

Equivalently,

$$\varepsilon_r = \sqrt{1 - \frac{\sum_{i=1}^{r}(q\sigma_i)^2}{\parallel X \parallel_F^2}}. \tag{8}$$

This quantity provides an explicit measure of the approximation error associated with the ordered representation in Equation (4).

This error measure refers specifically to the ordered representation in Equation (4). The actual truncated reconstructions used in the experiments are obtained from the reduced core tensor and the corresponding truncated factor matrices, as described below. The coefficients $q\sigma_i$ are ordered in non-increasing form as

$$q\sigma_1 \geq q\sigma_2 \geq \cdots \geq q\sigma_r \geq 0. \tag{9}$$

so that the first coefficients provide the dominant contribution to the progressive approximation.

For convenience, the ordered coefficients $q\sigma_i$ can be arranged along the main spatial diagonal of a compact 3D array S, that is,

$$S_{ijk} = \begin{cases} q\sigma_i, & \text{if } i = j = k, \\ 0, & \text{otherwise.} \end{cases} \quad (10)$$

This arrangement is used here as a simple structured way to organize the coefficients in the volumetric setting.

### 3.2 *Progressive Reconstruction and Coefficient Interpretation*

The quasi-singular coefficients $q\sigma_i$ are introduced here as practical ordered coefficients associated with the structured representation produced by the method. Their purpose is to guide truncation and progressive reconstruction in a simple way that supports structural interpretation of the volumetric data.

If $X_k$ denotes the approximation obtained for truncation level $k$, then increasing $k$ yields progressively more detailed reconstructions of the original volume. In this sense, the method provides a practical mechanism for hierarchical approximation, analogous to truncated matrix SVD and related multilinear truncation frameworks [1, 2].

To monitor this progressive behavior, cumulative coefficient-based curves can be computed. In this work, these curves are used as practical indicators of how the contribution of the retained quasi-singular coefficients grows as $k$ increases. This provides a simple way to compare reconstruction behavior across different truncation levels.

### 3.3 *Practical Features*

The proposed Structured 3D-SVD method has several practical features that make it suitable for volumetric reconstruction tasks:
- It operates directly in the spatial domain on the original volumetric data.
- It provides an ordered representation that supports progressive approximation and reconstruction.
- It is based on standard matrix factorizations along the tensor modes and is simple to implement.
- It allows reconstructions of different quality levels by varying the truncation parameter $k$.
- It offers a representation that supports structural interpretation of dominant volumetric patterns.

These features make the method useful for the compact representation, progressive visualization, and reconstruction of biological volumetric images.

### 3.4 *Interpretation of the Structured Components*

The proposed representation also admits a natural structural interpretation in the context of volumetric data. For each index $i$, the vectors $u^{(i)}$, $v^{(i)}$ and $w^{(i)}$ describe coordinated spatial patterns along the three tensor modes, while the corresponding quasi-singular coefficient $q\sigma_i$ quantifies the contribution of that triplet to the reconstructed volume. In this way, the representation provides an ordered description of the dominant spatial interactions present in the volumetric object.

From an application viewpoint, larger quasi-singular coefficients are associated with the most relevant global structures in the data, whereas smaller coefficients progressively contribute finer anatomical or morphological detail. Therefore, the ordered triplets $(\mathbf{u}^{(i)}, \mathbf{v}^{(i)}, \mathbf{w}^{(i)})$ can be understood as structured components that organize the reconstruction from coarse to fine information. This is particularly useful in biological volumetric imaging, where one is often interested in identifying the dominant overall shape first and then refining the reconstruction with increasingly detailed structural information.

Accordingly, the proposed framework provides interpretability in a structural and reconstruction-oriented sense: the coefficient ordering indicates the relative contribution of the components, and the corresponding mode vectors indicate how that information is distributed across the three spatial dimensions of the volume.

### 3.5 *Geometric Interpretation*

From a geometric point of view, the proposed method represents a three-dimensional volume as a combination of structured spatial components, each one obtained from three one-dimensional profiles and an associated coefficient.

Each term $q\sigma_i$ ($u^{(i)} \otimes v^{(i)} \otimes w^{(i)}$) can be interpreted as an ordered separable volumetric component. The vectors $u^{(i)}$, $v^{(i)}$ and $w^{(i)}$ describe coordinated spatial variation along the three principal directions of the volume, while $q\sigma_i$ indicates the relative contribution of that component within the ordered representation.

As the truncation level increases, the reconstruction captures an increasingly large portion of the volumetric structure and progressively approaches the original data.

### 3.6 *Direct Applications*

Owing to its simple formulation and progressive reconstruction capability, the proposed method is suitable for several practical tasks:
- Volume compression, by retaining the dominant coefficients for reconstruction.
- Progressive transmission, where the representation can be used to generate successively refined approximations.
- Partial reconstruction in resource-limited environments, where reduced representations are required.

The method could also be extended to related tasks such as denoising or structural inspection, although these applications are beyond the scope of the present work.

### 3.7 *Computational Structure*

A practical advantage of the proposed method is that it is computed directly in the spatial domain, without requiring frequency-domain transformations or alternative tensor signal representations. This makes it especially attractive when a direct volumetric reconstruction and a clear structural interpretation are desired.

In practice, the method computes a structured low-rank representation of the tensor and then generates truncated reconstructions for different values of $k$ from the reduced core tensor and the corresponding truncated factor matrices. Once the factor matrices and the reduced core have been obtained, approximations at different quality levels can be constructed efficiently.

### 3.8 *Algorithm for Structured 3D-SVD*

This section describes the computational procedure used to obtain the Structured 3D-SVD representation of a three-dimensional tensor $X \in R^{n1 \times n2 \times n3}$. The algorithm operates directly in the spatial domain and provides a low-rank representation that enables progressive reconstruction of the original volume.

The procedure is based on the matrix unfoldings of the tensor along the three modes and on the computation of matrix SVDs of these unfoldings. Rather than extracting rank-one terms sequentially from residual tensors, the method computes the representation up to a prescribed maximum rank $r$, from which truncated approximations for different values of $k \leq r$ can then be generated.

The computational cost of the proposed method is predominantly concentrated in the singular value decompositions of the three matrix unfoldings of the tensor. If standard matrix SVD is used, this leads to an approximate complexity of

$$O(n_1^2 n_2 n_3 + n_2^2 n_1 n_3 + n_3^2 n_1 n_2). \tag{11}$$

Once these decompositions have been computed, reconstructions for different quality levels can be obtained by truncating the representation and rebuilding the corresponding approximation, with a much smaller additional cost.

In comparison, Tucker decomposition typically combines an initialization stage of comparable order with an additional iterative refinement process. A simplified estimate of its complexity is

$$O(n_1^2 n_2 n_3 + n_2^2 n_1 n_3 + n_3^2 n_1 n_2 + T_T n_1 n_2 n_3 r). \tag{12}$$

where $T_T$ denotes the number of refinement iterations.

For CPD, a standard approximate complexity can be written as

$$O(T_C(n_1 n_2 n_3 r + (n_1 + n_2 + n_3)r^2 + r^3)). \tag{13}$$

where $T_C$ denotes the number of iterations. In practice, CPD is usually more sensitive to convergence and often requires a larger computational effort, which is consistent with the runtimes observed in the numerical experiments.

---

**Algorithm 1 Structured 3D-SVD Algorithm.**

---

Input: Tensor $X \in R^{n_1 \times n_2 \times n_3}$, maximum rank $r$

Output: Quasi-singular coefficients $q\sigma_i$, vectors $u^{(i)}, v^{(i)}, w^{(i)}$, and reconstructions $X_k$ for $k \leq r$

1: Form the modewise matrix unfoldings $X_{(1)}, X_{(2)}, X_{(3)}$ of X
2: Compute $X_{(m)} = \widetilde{U}_m \Sigma_m V_m^\top$, for $m = 1, 2, 3$
3: Set $U_m = U_{em}(:, 1:r)$, for $m = 1, 2, 3$
4: Compute $\mathcal{G} = \mathcal{X} \times_1 U_1^\top \times_2 U_2^\top \times_3 U_3^\top$
5: for $i = 1$ to $r$ do
6:    $u^{(i)} = U_1(:,i)$, $v^{(i)} = U_2(:,i)$, $w^{(i)} = U_3(:,i)$
7:    $q\sigma_i = G_{iii}$
8: end for
9: for each $k \leq r$ do
10:    $G_k = G(1:k, 1:k, 1:k)$
11:    $U_{m,k} = U_m(:, 1:k)$, for $m = 1, 2, 3$
12:    $X_k = G_k \times_1 U_{1,k} \times_2 U_{2,k} \times_3 U_{3,k}$
13: end for
14: return $\{q\sigma_i, u^{(i)}, v^{(i)}, w^{(i)}\}_{i=1}^r$ and $X_k$

---

The matrices $X_{(1)}, X_{(2)}, X_{(3)}$ denote the modewise unfoldings of the tensor X, obtained by rearranging its entries into matrix form along modes 1, 2, and 3, respectively.

For each truncation level $k$, the reconstruction is obtained from the truncated core tensor and the corresponding truncated factor matrices, rather than from a purely diagonal expansion.

In this implementation, the quasi-singular coefficients $q\sigma_i$ are defined as the ordered main spatial diagonal entries of the reduced core tensor, while the associated vectors are given by the corresponding basis vectors of the truncated mode matrices. These coefficients are used as practical indicators for truncation and progressive reconstruction.

## 4 Experimental Evaluation and Results

To evaluate the performance of the Structured 3D-SVD algorithm on volumetric biological data, we carried out a set of experiments on two 3D datasets following the procedure described in Algorithm 1: a full-volume scan of a fish and a full-volume scan of a brain. The study has two main objectives: first, to determine whether the method can capture the main volumetric structure at reduced truncation levels, and second, to evaluate how the reconstruction quality improves as the truncation level increases. The method provides a representation based on ordered coefficients that enables a progressive analysis of reconstruction accuracy as the truncation level increases.

### 4.1 *Evaluation Metrics*

To quantify the obtained results, four evaluation measures were used:

(a) Peak signal-to-noise ratio (PSNR):

$$\text{PSNR} = 10 \cdot \log_{10}\left(\frac{I_{\max}^2}{\text{MSE}}\right), \quad (14)$$

where $I_{\max}$ denotes the maximum intensity value in the volume. This metric is widely used in image processing to assess reconstruction quality, with higher PSNR values indicating better reconstructions.

(b) Mean squared error (MSE):

$$\text{MSE} = \frac{1}{n_1 n_2 n_3} \sum_{i,j,k} \left(X_{ijk} - \hat{X}_{ijk}\right)^2. \quad (15)$$

This metric measures the average discrepancy between the original volume and its reconstructed version.

(c) Relative reconstruction error (RelErr):

$$\text{RelErr} = \frac{\|X - \hat{X}\|_F}{\|X\|_F}, \quad (16)$$

where X denotes the original tensor, X̂ the reconstructed tensor, and $\|\cdot\|_F$ the Frobenius norm.

(d) Percentage of energy retained (PER):

$$\text{PER}(k) = \frac{\sum_{i=1}^{k}(q\sigma_i)^2}{\sum_{i=1}^{r_{\max}}(q\sigma_i)^2}. \quad (17)$$

In this work, PER is used as a cumulative coefficient-based indicator associated with the ordered quasi-singular coefficients. It reflects the relative accumulated energy of these coefficients as the truncation level $k$ increases, and it is used here as a practical guide for progressive reconstruction analysis.

These measures are complemented by visual reconstructions and quantitative comparisons. Specific examples, visualizations, and a detailed analysis of the results are presented below.

4.2  *Experimental Design and Setup*

The implementation was developed in Python 3.11, using the NumPy and SciPy libraries. All experiments were conducted on a workstation with an Intel Core i7 CPU, 16 GB RAM, and no GPU acceleration. We used publicly available 3D biological imaging data obtained from the MorphoSource repository, which provides high-resolution scans of animal specimens. In our experiments, we selected the fish and brain volumetric datasets in order to evaluate the method on real anatomical structures.

All volumetric data were converted to real-valued third-order tensors and normalized to the range [0,1] before decomposition. No denoising, filtering, or other preprocessing operations were applied beyond cropping/rescaling to the target tensor sizes used in the experiments. Slice selection for visual comparison was fixed in advance and kept identical across methods in order to ensure a consistent qualitative evaluation. Specifically, the reported visual reconstructions correspond to slices 80, 100, and 120 for both datasets.

For the baseline methods, Tucker decomposition was computed with multilinear ranks $(k,k,k)$ and standard SVD-based initialization, while CPD was computed with rank $k$, random initialization, maximum number of iterations 300, tolerance $10^{-6}$, and factor normalization enabled. For Structured 3D-SVD, the representation was precomputed up to the maximum rank $r$, and reconstructions were then generated for the corresponding truncation levels $k \leq r$. All methods were evaluated on the same normalized tensors and at the same reported truncation levels in order to ensure a fair comparison.

Since CPD depends on random initialization, it was repeated 10 times using different seeds, and the reported CPD values are presented as the means and 95% confidence intervals. Under the reported settings, Tucker and Structured 3D-SVD are deterministic, and so repeated runs yield identical results up to numerical precision.

For CPD and for Tucker, when iterative refinement is involved, the reported iteration limits are used only as numerical optimization settings associated with the convergence of the decomposition algorithm. In the present work, they are not interpreted as explicit mechanisms for controlling overfitting, since the comparison is focused on reconstruction behavior under fixed rank/truncation settings rather than on predictive generalization.

4.3  *Experimental Results*

For the fish dataset, the data were represented as 3D tensors $X \in R^{80 \times 100 \times 180}$. For computational convenience, the original volumes were cropped and rescaled to these target tensor dimensions before the compared decompositions were applied.

The Structured 3D-SVD decomposition was computed up to $r = 80$. For the visual reconstruction analysis, only the truncation levels $k = 20$ and $k = 50$ were considered.

Figure 1 displays the original slices together with the corresponding reconstructions for $k = 20$ and $k = 50$, using slices 80, 100, and 120. The reconstruction quality improves as $k$ increases, with the main anatomical structures becoming progressively better preserved.

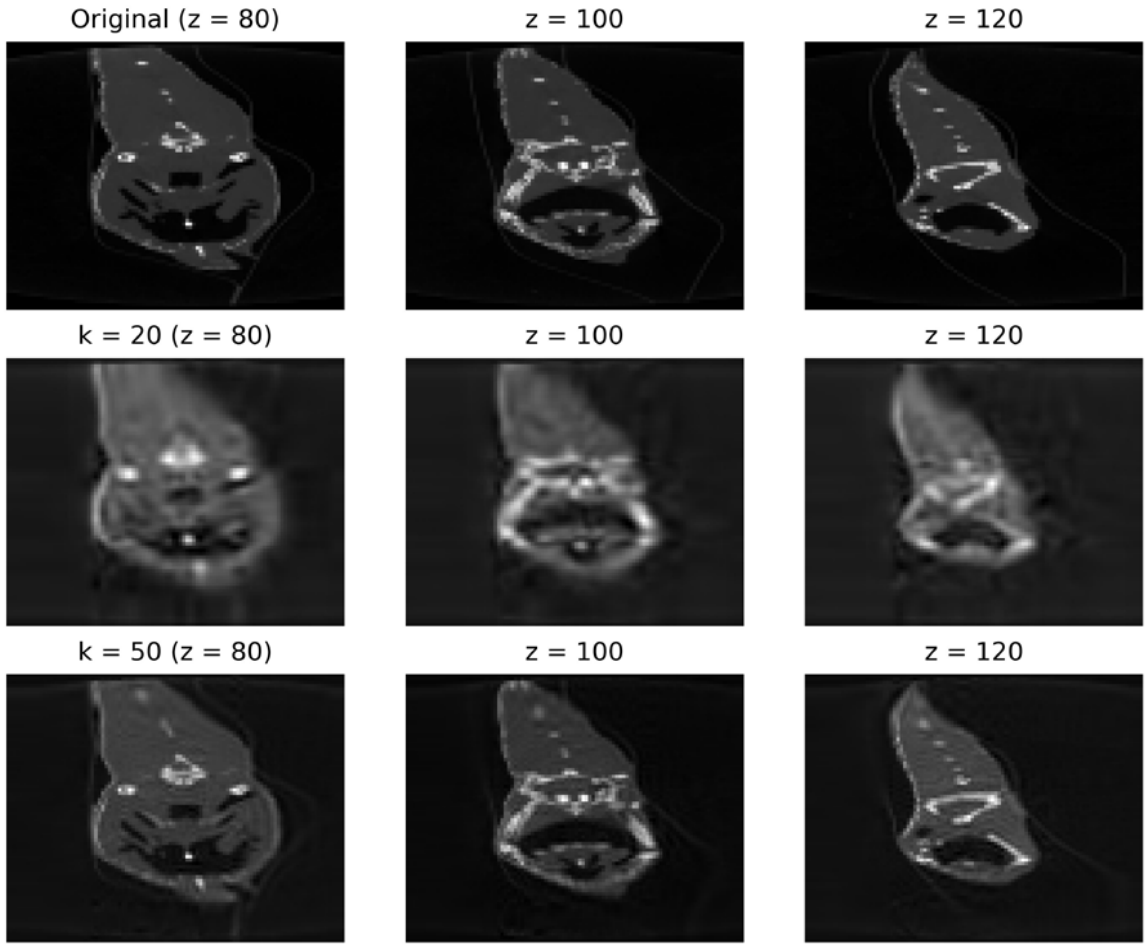

**Figure 1**: Reconstruction of slices 80, 100, and 120 from the fish volumetric dataset for truncation levels $k = 20$ and $k = 50$.

To provide a baseline comparison, the proposed method was compared with Tucker and CP tensor decompositions on the same datasets and at the same truncation levels.

Table 2 shows that Tucker and Structured 3D-SVD provide very similar reconstruction quality for all tested truncation levels, while Structured 3D-SVD requires consistently lower computation times. For instance, at $k = 20$, Tucker reaches 29.60 dB, whereas Structured 3D-SVD attains 29.43 dB with a shorter execution time (0.486 s versus 0.763 s). A similar trend is observed at $k = 30$ and $k = 50$, where the proposed method remains close to Tucker in PSNR, MSE, and relative error while requiring less computation time.

**Table 2:** Comparison of reconstruction quality and computation time for Tucker, CPD, and Structured 3D-SVD decompositions on the fish volumetric dataset.

| Method | $k$ | PSNR (dB) | MSE | RelErr | Time (s) |
|---|---|---|---|---|---|
| Tucker | 20 | 29.60 | $1.096 \times 10^{-3}$ | 0.238 | 0.763 |
| Tucker | 30 | 31.78 | $6.633 \times 10^{-4}$ | 0.185 | 0.962 |
| Tucker | 50 | 35.15 | $3.048 \times 10^{-4}$ | 0.125 | 1.195 |
| CPD | 20 | 28.05 | $1.564 \times 10^{-3}$ | 0.284 | 3.270 |
| CPD | 30 | 28.82 | $1.310 \times 10^{-3}$ | 0.260 | 4.536 |
| CPD | 50 | 29.77 | $1.052 \times 10^{-3}$ | 0.233 | 6.679 |
| Structured 3D-SVD | 20 | 29.43 | $1.139 \times 10^{-3}$ | 0.242 | 0.486 |
| Structured 3D-SVD | 30 | 31.59 | $6.919 \times 10^{-4}$ | 0.189 | 0.526 |
| Structured 3D-SVD | 50 | 35.08 | $3.103 \times 10^{-4}$ | 0.126 | 0.536 |

By contrast, CPD gives lower reconstruction quality at all tested ranks and requires substantially longer runtimes. These results indicate that the proposed Structured 3D-SVD provides a favorable compromise between accuracy and efficiency, achieving performance close to Tucker with lower computation times, while remaining substantially faster than CPD.

For CPD, additional runs with different random initializations were performed. The corresponding mean values and 95% confidence intervals for the fish dataset are reported separately in Table 3.

**Table 3:** CPD results on the fish dataset reported as mean ± 95% confidence intervals over 10 runs with different random initializations.

| $k$ | PSNR (dB) | MSE | RelErr | Time (s) |
|---|---|---|---|---|
| 20 | 28.05 ± 0.02 | $1.564 \times 10^{-3} \pm 7.71 \times 10^{-6}$ | $0.284 \pm 7.11 \times 10^{-4}$ | 3.270 ± 0.0357 |
| 30 | 28.82 ± 0.02 | $1.310 \times 10^{-3} \pm 7.44 \times 10^{-6}$ | $0.260 \pm 7.41 \times 10^{-4}$ | 4.536 ± 0.0519 |
| 50 | 29.77 ± 0.02 | $1.052 \times 10^{-3} \pm 5.90 \times 10^{-6}$ | $0.233 \pm 6.56 \times 10^{-4}$ | 6.679 ± 0.0420 |

Table 4 provides a more detailed view of the progressive reconstruction behavior on the fish dataset. As the truncation level $k$ increases, both the PSNR and the PER improve steadily. In particular, the PSNR rises from 29.43 dB at $k = 20$ to 35.08 dB at $k = 50$, while the PER increases from 94.60% to 98.39%. This con rms that relatively low truncation levels already preserve the main volumetric structures, whereas larger values of $k$ lead to increasingly refined reconstructions. The progressive behavior observed in the reconstructions is consistent with this interpretation: the rst components capture the main volumetric organization of the biological object, while additional components re ne anatomical or morphological detail as the truncation level increases.

**Table 4**: Reconstruction quality metrics for different values of rank $k$ on the fish volumetric dataset.

| $k$ | PSNR (dB) | PER (%) |
|---|---|---|
| 10 | 27.58 | 90.97 |
| 20 | 29.43 | 94.60 |
| 30 | 31.59 | 96.42 |
| 40 | 33.30 | 97.58 |
| 50 | 35.08 | 98.39 |
| 60 | 37.09 | 98.99 |
| 70 | 39.24 | 99.38 |
| 80 | 41.10 | 99.60 |

Table 5 con rms the behavior observed on the fish dataset. Structured 3D-SVD provides reconstruction quality very close to that of Tucker while requiring less computation time in all tested cases. By contrast, CPD again gives lower reconstruction quality and much longer runtimes. Overall, the results on the brain dataset support the same conclusion: Structured 3D-SVD achieves a favorable compromise between reconstruction accuracy and computational efficiency.

**Table 5:** Comparison of reconstruction quality and computation time for Tucker, CPD, and Structured 3D-SVD decompositions on the brain volumetric dataset.

| Method | $k$ | PSNR (dB) | MSE | RelErr | Time (s) |
|---|---|---|---|---|---|
| Tucker | 20 | 23.92 | $4.051 \times 10^{-3}$ | 0.224 | 3.935 |
| Tucker | 30 | 26.17 | $2.411 \times 10^{-3}$ | 0.173 | 3.965 |
| Tucker | 50 | 30.39 | $9.139 \times 10^{-4}$ | 0.106 | 3.984 |
| CPD | 20 | 21.76 | $6.665 \times 10^{-3}$ | 0.288 | 10.775 |
| CPD | 30 | 22.79 | $5.253 \times 10^{-3}$ | 0.256 | 13.268 |
| CPD | 50 | 24.01 | $3.967 \times 10^{-3}$ | 0.222 | 19.830 |
| Structured 3D-SVD | 20 | 23.68 | $4.284 \times 10^{-3}$ | 0.231 | 2.569 |
| Structured 3D-SVD | 30 | 26.03 | $2.490 \times 10^{-3}$ | 0.176 | 2.571 |
| Structured 3D-SVD | 50 | 30.33 | $9.251 \times 10^{-4}$ | 0.107 | 2.578 |

The same repeated run analysis was also carried out for CPD on the brain dataset. The resulting mean values together with the corresponding 95% confidence intervals are reported separately in Table 6.

**Table 6**: CPD results on the brain dataset, reported as means and 95% condidence intervals over 10 runs with different random initializations.

| $k$ | PSNR (dB) | MSE | RelErr | Time (s) |
|---|---|---|---|---|
| 20 | 21.76 ± 0.06 | 6.665×10$^{-3}$ ± 9.27×10$^{-5}$ | 0.288 ± 2.01×10$^{-3}$ | 10.775 ± 0.0387 |
| 30 | 22.79 ± 0.01 | 5.253×10$^{-3}$ ± 1.49×10$^{-5}$ | 0.256 ± 3.62×10$^{-4}$ | 13.268 ± 0.0384 |
| 50 | 24.01 ± 0.02 | 3.967×10$^{-3}$ ± 1.64×10$^{-5}$ | 0.222 ± 4.59×10$^{-4}$ | 19.830 ± 0.0863 |

For the brain dataset, the tensor size was $\mathbb{R}^{180\times210\times180}$. As in the fish dataset, Figure 2 shows the visual reconstructions for slices 80, 100, and 120 at truncation levels $k = 20$ and $k = 50$. The reconstruction quality also improves as $k$ increases, with the anatomical structures becoming more clearly preserved.

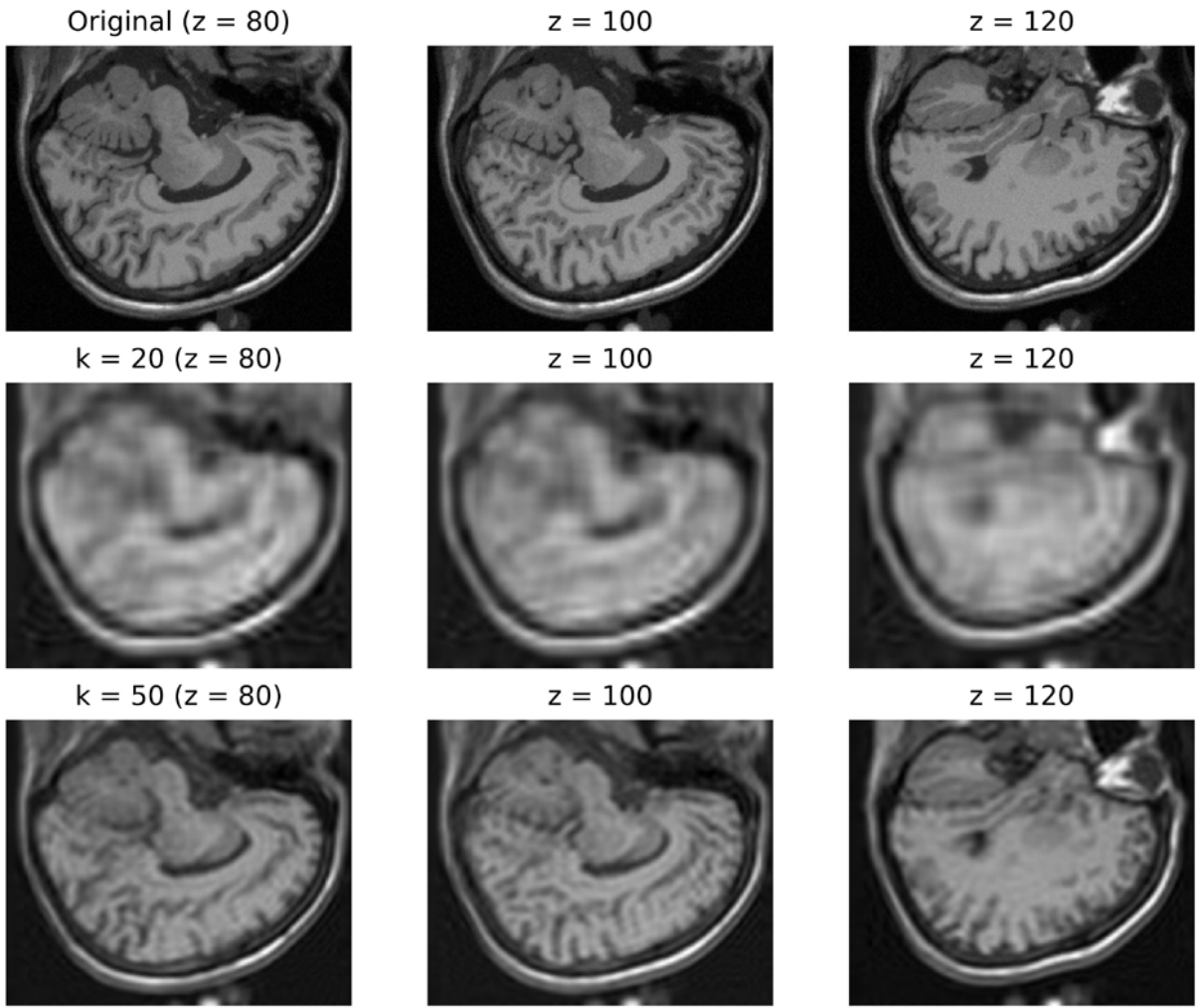

**Figure 2**: Reconstruction of slices 80, 100, and 120 from the brain volumetric dataset for truncation levels $k = 20$ and $k = 50$.

Figure 3 shows the accumulated PER as a function of the truncation level $k$ for the fish and brain volumetric datasets. In both cases, the curves increase rapidly for small values of $k$ and then gradually approach saturation, indicating that the rst truncation levels account for most of the retained contribution in the ordered representation.

The dashed horizontal line marks the 99% PER threshold, which is reached at $k = 54$ for the fish dataset and at $k = 61$ for the brain dataset. This behavior is consistent with the progressive reconstruction results, showing that relatively low truncation levels already preserve most of the relevant volumetric information.

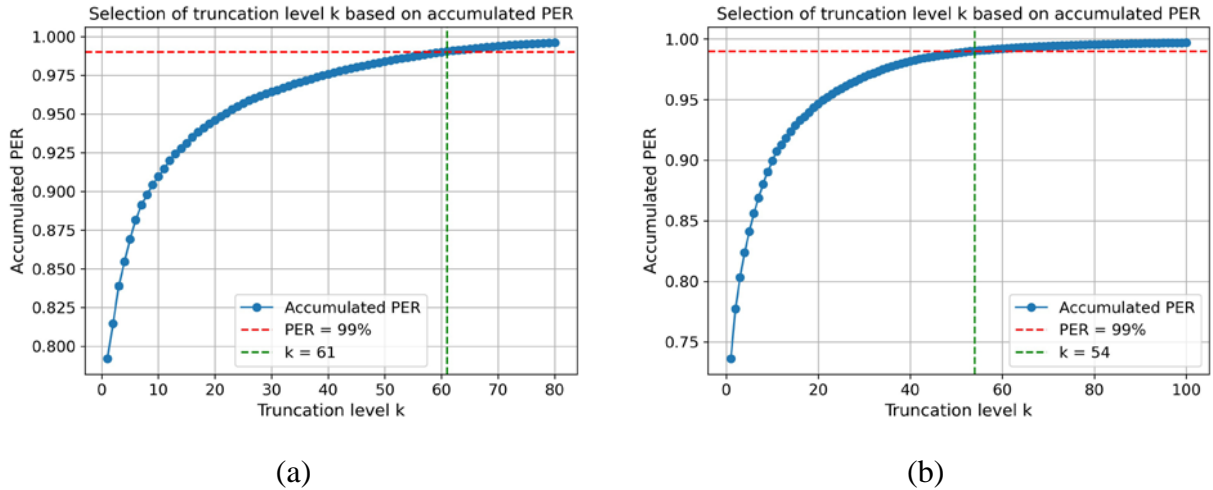

(a)  (b)

**Figure 3:** Accumulated PER as a function of the truncation level $k$ for (a) the fish volumetric dataset and (b) the brain volumetric dataset.

Figure 4 shows the evolution of the PSNR as a function of the truncation level $k$ for the fish and brain volumetric datasets. In both cases, the PSNR increases steadily as the truncation level increases, con rming the progressive reconstruction capability of the proposed method.

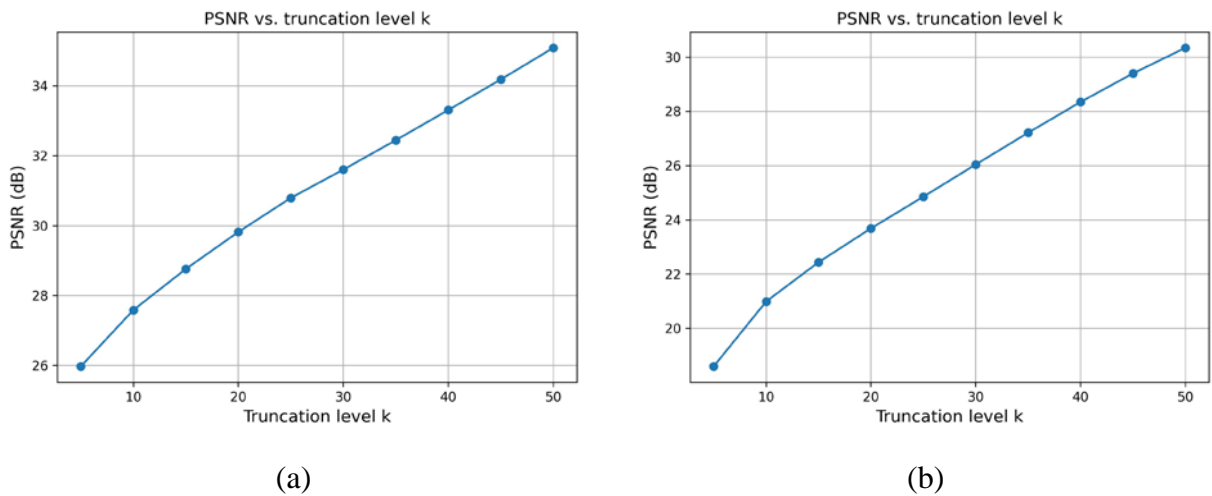

(a)  (b)

**Figure 4**: PSNR vs. truncation level $k$ for (a) the fish volumetric dataset and (b) the brain volumetric dataset.

The most signi cant improvement is observed at lower values of $k$, while the gain remains positive and regular up to $k = 50$. The fish dataset reaches higher PSNR values than the brain dataset for the same truncation levels, which suggests that the brain volume is more challenging to approximate at the same truncation levels.

### 4.4 *Sensitivity Analysis and Practical Selection of r and k*

The sensitivity of the proposed method with respect to the truncation level $k$ can be analyzed from the PER and PSNR curves shown in Figures 3 and 4. In both datasets, the PER increases rapidly for small values of $k$ and then gradually approaches saturation, while the PSNR improves steadily as $k$ increases. This behavior indicates that the method is not overly sensitive to small variations of $k$ once the main volumetric structures have been captured and that progressively larger values of $k$ mainly contribute to ner reconstruction details.

From a practical viewpoint, the PER can be used as a simple criterion to select the truncation level $k$. In particular, $k$ may be chosen as the smallest value such that PER($k$) reaches a prescribed threshold, depending on the desired trade-off between compactness and reconstruction quality. In the present experiments, the 99% PER threshold is reached at $k = 54$ for the fish dataset and at $k = 61$ for the brain dataset, which provides a practical reference for selecting truncation levels that preserve most of the relevant volumetric information.

A complementary sensitivity analysis with respect to the maximum precomputed rank $r$ was also carried out. For this purpose, the truncation levels $k = \{20, 30, 50\}$ were evaluated, while the maximum rank was varied as $r = \{50, 60, 80\}$ for the fish dataset and $r = \{50, 80, 100\}$ for the brain dataset. For both datasets, the PSNR, PER, and relative error values remained unchanged for fixed $k$ whenever $r \geq k$. This indicates that, in the present implementation, $k$ is the e ective reconstruction parameter, whereas $r$ acts as an upper bound on the reduced representation computed once in advance. Therefore, in practice, $r$ should be selected to be at least as large as the maximum truncation level of interest.

A qualitative comparison of sensitivity across the considered methods can also be drawn from the reported experiments. Structured 3D-SVD shows stable progressive behavior with respect to the truncation level $k$, while remaining invariant with respect to the maximum precomputed rank $r$ whenever $r \geq k$. Tucker decomposition is deterministic under the reported settings and exhibits the expected dependence on the selected multilinear rank. CPD, in contrast, is sensitive not only to the chosen rank but also to random initialization, which is re ected in the repeated run analysis reported for both datasets. In this sense, the proposed Structured 3D-SVD framework combines progressive truncation sensitivity with deterministic behavior, whereas CPD introduces an additional source of variability associated with optimization initialization.

### 4.5 *Relation to Other Tensor and Learning-Based Approaches*

Beyond Tucker and CPD, other families of methods have also been considered in the literature for multidimensional data representation and reconstruction. In particular, tensor-train and tensor-ring decompositions are based on different low-rank representation principles and are often considered in settings involving higher-order tensors, while deep learning-based approaches are typically oriented toward data-driven and task-specific optimization when large training datasets are available.

In comparison, the proposed Structured 3D-SVD framework is aimed at a different tradeo . It is designed for third-order volumetric data, operates directly in the spatial domain, and does not require training or iterative model tting beyond the initial factor extraction. Its main practical advantages are simplicity, interpretability, and the ability to support progressive reconstruction through ordered quasi-singular coefficients.

At the same time, the proposed method is not intended to replace tensor-network or deep learning-based approaches in all scenarios. Methods such as tensor train or tensor ring are based on different low-rank representation principles and are often considered in settings involving higher-order tensors, while deep learning-based models are typically oriented toward data-driven optimization when sufficient training data are available. Therefore, the present approach is best understood as a simple and interpretable framework for progressive truncation and reconstruction of 3D volumetric data, rather than as a universal alternative to all modern tensor or learningbased methods.

### 4.6 *Limitations and Practical Scope*

The proposed Structured 3D-SVD framework is intended for a specific range of applications and should be interpreted within that scope. A main limitation of the proposed method is that its current formulation is specifically designed for third-order volumetric data in the spatial domain, and so its present scope does not extend to arbitrary higher-order tensor settings. In addition, although the method provides an ordered and interpretable truncation mechanism, it does not introduce a canonical tensor spectrum with the same theoretical status as matrix singular values. The experimental validation in this work is based on biological volumetric datasets, and so the present results should be interpreted within that scope. Extending the evaluation to more complex or higher-dimensional datasets constitutes a natural direction for future work.

The method is particularly suited to scenarios in which a simple, training-free, and progressively reconstructible representation of 3D volumetric data is required. This includes cases where interpretability, moderate computational cost, and direct operation in the spatial domain are especially relevant. In other application settings, tensor-network or data-driven approaches may be considered depending on the modeling objectives, the structure of the data, and the availability of training resources. In real-world systems, the proposed approach is therefore best viewed as a practical reconstruction and truncation framework for structured 3D volumetric data, particularly when progressive approximation and straightforward implementation are relevant requirements.

From a practical perspective, the proposed method is particularly suitable when (i) the data are third-order volumetric tensors, (ii) a training-free and deterministic reconstruction framework is preferred, (iii) progressive truncation and reconstruction are relevant, and (iv) structural interpretability in the spatial domain is of interest.

## 5 Conclusions

This work presented Structured 3D-SVD as a practical method for the compression, reconstruction, and analysis of volumetric biological data. Inspired by the logic of matrix SVD, the method operates directly in the spatial domain and provides a representation of 3D volumes that supports progressive reconstruction.

The experimental results on the fish and brain volumetric datasets showed that the proposed method achieves reconstruction quality very close to that of Tucker decomposition while requiring shorter computation times. On both datasets, it also outperformed CPD in terms of both reconstruction quality and runtime. The progressive reconstruction analysis further showed that relatively low truncation levels are sufficient to preserve the main volumetric structures, while higher truncation levels lead to more detailed reconstructions.

These results support the use of Structured 3D-SVD as a practical framework for compact volumetric representation, progressive truncation, reconstruction, and visualization of biological images, with quasi-singular coefficients acting as ordered indicators within the proposed representation, especially in applications where both efficiency and progressive structural interpretation are important.

Possible extensions include adapting the method to anisotropic or irregular sampling, incorporating application-dependent constraints into the representation, and developing GPU implementations for larger biomedical datasets. It would also be of interest to explore integrating Structured 3D-SVD into machine learning workflows involving volumetric data, such as classification or registration.